\documentclass[twocolumn,fp]{jpsj3}

\usepackage{txfonts}

\usepackage{amsmath}
\usepackage{amsfonts}
\usepackage{bm}
\usepackage{graphicx}
\usepackage{color}

\title{Machine learning technique to find quantum many-body ground states
of bosons on a lattice
}

\author{Hiroki Saito and Masaya Kato}
\inst{Department of Engineering Science, University of
Electro-Communications, Tokyo 182-8585, Japan}

\abst{
We develop a variational method to obtain many-body ground states of the
Bose-Hubbard model using feedforward artificial neural networks.
A fully-connected network with a single hidden layer works better than
a fully-connected network with multiple hidden layers, and a multi-layer
convolutional network is more efficient than a fully-connected network.
AdaGrad and Adam are optimization methods that work well.
Moreover, we show that many-body ground states with different numbers
of atoms can be generated by a single network.
}

\begin{document}
\maketitle

\section{Introduction}

Recent developments in the field of artificial neural networks (ANNs), in
combination with high-performance computers, have dramatically increased
the ability of artificial intelligence.
Techniques used in ANNs and machine learning have been applied to a wide
variety of fields not only in engineering, but also in science, including
physics research.

Pattern recognition is an important application of machine learning.
By training an ANN with a large amount of data, e.g., many sample
pictures, some features are extracted from the image data, and the ANN
becomes able to classify the pictures.
In physics, this ability of ANNs can be used to classify numerically or
experimentally obtained data, which are sometimes complicated and cannot
be identified by humans.
Trained ANNs can discriminate different phases of numerically or
analytically obtained many-body
states~\cite{Ohtsuki,Carra,Nieu,Zhang,Ohtsuki07,Broecker2,Tanaka,ChngX,Broecker,Chng,PZhang,Mano}.
A similar approach was developed for the analysis of experimental
data~\cite{Ovch} and telescope images~\cite{Hezaveh}.

In the above example of the classification of pictures, the memory size of
the ANN is much smaller than the total size of the image data used for
training.
Nevertheless, the ANN acquires features of sample pictures and can even
reproduce similar pictures.
This implies that ANNs can efficiently encode and store the features of
large amount of data, which is also applicable to physics problems.
Many-body quantum states in a large Hilbert space can be efficiently
stored in ANNs~\cite{Carleo,Deng,Chen,Gao,YHuang,Torlai07,Cai,Schmitt}.
Thermal fluctuations can also be learned by ANNs; i.e., ANNs trained by
Monte Carlo samples at finite temperature can reproduce thermodynamic
properties~\cite{Torlai,Morning}.
Such trained ANNs can be used for efficient Monte Carlo
updates~\cite{Huang,Wang}.
Complicated functions of many variables can be stored in ANNs, which has
been used for efficient simulation of molecular dynamics~\cite{Behler}.
Quantum error corrections are also possible using ANNs~\cite{TorlaiL}.

Recently, a method to solve quantum many-body problems using ANNs was
proposed~\cite{Carleo}.
It was demonstrated that the ground states and time evolutions of the
quantum Ising and Heisenberg models can be obtained using ANNs.
In this method, a quantum state is represented by a restricted Boltzmann
machine, which consists of input and hidden units.
When a spin configuration $\uparrow \downarrow \cdots$ is set to the input
units, the corresponding wave function $\psi(\uparrow \downarrow \cdots)$
is obtained from the ANN.
The internal parameters of the network are optimized in such a way that
the wave functions produced by the ANN satisfy the desired properties,
e.g., energy minimization.

Motivated by Ref.~\citen{Carleo}, a method to treat the Bose-Hubbard model
was proposed in Ref.~\citen{Letter}, where the feedforward network
was used instead of the restricted Boltzmann machine.
It was shown that the many-body ground state obtained by the method of ANN
agrees very well with that obtained by exact diagonalization, even when
the number of bases in the Hilbert space is much larger than the number of
network parameters.
This implies that the method proposed in Ref.~\citen{Carleo} is applicable
to a broad class of quantum many-body problems and that ANNs with machine
learning are powerful tools to explore quantum many-body physics.

The present paper provides the extended results of the Letter in
Ref.~\citen{Letter}, in which a fully-connected network with a single
hidden layer was only used with a simple steepest-descent method to
optimize the network.
In the present paper, we examine fully-connected networks and
convolutional networks with multiple hidden layers.
We show that a fully-connected network having a single hidden layer with
sufficient units yields better ground states than that with multiple
hidden layers.
On the other hand, a convolutional network with multiple hidden layers is
more efficient than a convolutional network with a single hidden layer or
a fully-connected network.
With respect to methods for optimizing ANNs, AdaGrad~\cite{Duchi} and
Adam~\cite{Kingma} allow faster convergence than the simple
steepest-descent method.
We also show that many-body ground states with different numbers of atoms
can be generated by a single ANN that has been multiply optimized for
these numbers of atoms.
Even when an ANN is optimized for a specific number of atoms, ground
states with other numbers of atoms can be extrapolated approximately.

The remainder of the present paper is organized as follows.
Section~\ref{s:method} explains the method to obtain the quantum many-body
ground state using an ANN.
Section~\ref{s:bose} provides the results of numerical calculations.
Section~\ref{s:conc} presents the conclusions of the present study.

\section{Method}
\label{s:method}

\subsection{Network architectures}

\begin{figure}
\begin{center}
\includegraphics[width=8.5cm]{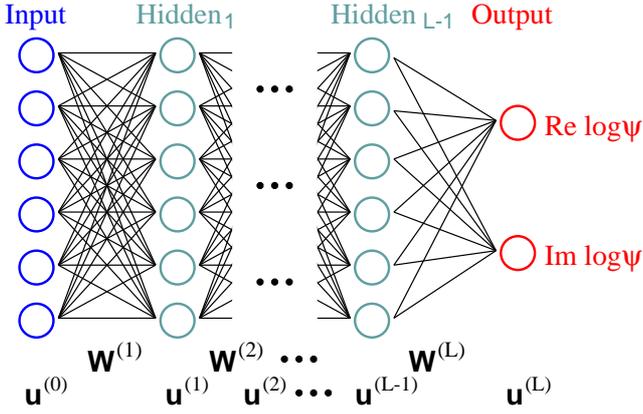}
\end{center}
\caption{
Schematic diagram of the feedforward network.
The input and output units are expressed as $\bm{u}^{(0)}$ and
$\bm{u}^{(L)}$, respectively, and $\bm{u}^{(1)}, \cdots, \bm{u}^{(L-1)}$  
are hidden units.
Given the values in the input units, the feedforward propagation generates
the values in the output units, from which the wave function is calculated
as $\psi = \exp[u_1^{L} + i u_2^{L}]$.
Some or all of the units in the $(n-1)$th layer are connected to those in
the $n$th layer through the weights $\bm{W}^{(n)}$.
}
\label{f:schematic}
\end{figure}
We use the feedforward network illustrated in Fig.~\ref{f:schematic}.
The ANN consists of input, hidden, and output layers.
The units in the input and output layers are denoted by $\bm{u}^{(0)}$ and
$\bm{u}^{(L)}$, respectively, and those in the hidden layers are denoted
by $\bm{u}^{(1)}$, $\bm{u}^{(2)}$, $\cdots$, $\bm{u}^{(L-1)}$.
The number of units in the $n$th layer is written as $N_n$, and the number
of units in the output layer is fixed as $N_L = 2$.
Some or all of the units in $\bm{u}^{(n-1)}$ are connected to those in
$\bm{u}^{(n)}$ through the weights $\bm{W}^{(n)}$.

In the present paper, we examine two types of feedforward neural networks.
The first is a fully-connected neural network, in which each unit in
the $(n-1)$th layer is connected to all of the units in the $n$th layer.
The network parameters are the weights $\bm{W}^{(n)}$ and biases
$\bm{b}^{(n)}$, which are real $N_n \times N_{n-1}$ matrices and
$N_n$-component vectors, respectively.
First, we set values to the input units $u^{(0)}_i$, which are transferred
to the next layer as
\begin{equation} \label{ff1}
u^{(1)}_j = \sum_{i=1}^{N_0} W^{(1)}_{ij} u^{(0)}_i + b^{(1)}_j.
\end{equation}
In the hidden layers, an activation function $f$ is applied as
\begin{equation} \label{ff2}
u^{(n+1)}_j = \sum_{i=1}^{N_n} W^{(n+1)}_{ij} f(u^{(n)}_i) + b^{(n+1)}_j
\qquad (n \geq 1),
\end{equation}
where $f$ must be a nonlinear function, and here we adopt
\begin{equation}
f(x) = \tanh x.
\end{equation}
In the final layer, the bias is absent, $\bm{b}^{(L)} = \bm{0}$, because
the wave function is only multiplied by an overall factor (see
Eq.~(\ref{psi})).
The total number of network parameters for the fully-connected network
is then
\begin{equation} \label{NFC}
N_{\rm FC} = \sum_{n=1}^{L-1} N_n (N_{n-1} + 1) + 2 N_{L-1}.
\end{equation}

The second one is the convolutional neural network, in which the weights
$\bm{W}^{(n)}$ act as local filters and the units $\bm{u}^{(n)}$ consist
of multiple channels.
The values set in the input units $\bm{u}^{(0)}$ are transferred to the
next layer as
\begin{equation} \label{conv1}
u^{(1)}_{j,k} = \sum_{p=0}^{F_1-1} W^{(1)}_{pk} u^{(0)}_{j+p} +
b^{(1)}_k,
\end{equation}
where the index $k$ denotes the channel, and $F_1$ is the size of the
filter $W^{(1)}_{pk}$ for each channel.
In Eq.~(\ref{conv1}), $u^{(1)}_{j,k}$ is generated only from
$u^{(0)}_j$, $u^{(0)}_{j+1}$, $\cdots$, and $u^{(0)}_{j+F_1-1}$, and thus
the local feature in the input units is captured by each filter and
transferred into each channel.
The number of units in $\bm{u}^{(1)}$ is $N_1 = N_0 C_1$, where $C_n$ is
the number of channels in the $n$th layer.
The subsequent convolutional layers have the form,
\begin{equation} \label{conv2}
u^{(n)}_{j,m} = \sum_{k=1}^{C_{n-1}} \sum_{p=0}^{F_n-1} W^{(n)}_{pmk}
f(u^{(n-1)}_{j+p,k}) + b^{(n)}_m,
\end{equation}
where all of the $C_{n-1}$ channels in $\bm{u}^{(n-1)}$ are filtered and
summed to generate each output channel.
The numbers of units in $\bm{u}^{n-1}$ and $\bm{u}^n$ are thus $N_0
C_{n-1}$ and $N_0 C_n$, respectively.
Finally, the units $\bm{u}^{(L-1)}$ produced by the convolution layers
are fully-connected to the output units as
\begin{equation} \label{conv3}
u_{j}^{(L)} = \sum_{m=1}^{C_{L-1}} \sum_{i=1}^{N_0} W_{imj}^{(L)}
f(u_{i,m}^{(L-1)}),
\end{equation}
where $j = 1$, 2.
The total number of network parameters is thus
\begin{equation} \label{NConv}
N_{\rm conv} = (F_1 + 1) C_1 + \sum_{n=2}^{L-1} (F_n C_{n-1} + 1) C_n
+ 2 N_0 C_{L-1}.
\end{equation}
In the terminology of ANN, this network consists of multiple convolution
layers with a unit stride and no pooling layers, followed by a
fully-connected layer.

\subsection{Quantum many-body states}

A quantum many-body state of bosons on a lattice is expressed by the
feedforward ANN as follows.
An arbitrary state can be expanded by the Fock states as
\begin{equation} \label{Psi}
|\Psi \rangle = \sum_{\bm{n}} \psi(\bm{n}) |\bm{n} \rangle,
\end{equation}
where $\bm{n} = (n_1, n_2, \cdots, n_M)$ represents a particle
distribution on the lattice sites, with $M$ being the number of sites.
For the total number of particles $N$, the number of Fock state bases,
i.e., the number of $\bm{n}$ satisfying $\sum_{i=1}^M n_i = N$ is
\begin{equation} \label{nfock}
N_{\rm Fock} = \frac{(N + M - 1)!}{N! (M - 1)!}.
\end{equation}

The number of input units is taken to be $N_0 = M$.
We set the input units as
\begin{equation}
u_i^{(0)} = n_i - N / M \qquad (i = 1, \cdots, M),
\end{equation}
and the feedforward propagation in Eqs.~(\ref{ff1}) and (\ref{ff2})
[or Eqs.~(\ref{conv1}) and (\ref{conv2})] is then performed.
The wave function $\psi(\bm{n})$ is calculated from the output units
$\bm{u}^{(L)}$ as
\begin{equation} \label{psi}
\psi(\bm{n}) = \exp[ u^{(L)}_1 + i u^{(L)}_2 ].
\end{equation}
Although the ground-state wave function of bosons can be taken to be real
and positive, we include the phase $u^{(L)}_2$ in Eq.~(\ref{psi}) for
future use.
By calculating $\psi(\bm{n})$ for all possible $\bm{n}$, we can construct
the many-body state in Eq.~(\ref{Psi}).
The information of the many-body quantum state is therefore stored in the
network parameters $\bm{W}^{(n)}$ and $\bm{b}^{(n)}$.
Our aim is to optimize the network parameters so that the corresponding
many-body quantum state is as close to the ground state as possible.

Expectation values of quantities are calculated by the Monte Carlo method
with Metropolis sampling.
If we adopt the trial $\bm{n}_1 \rightarrow \bm{n}_2$ with the probability
${\rm min}[1, |\psi(\bm{n}_2) / \psi(\bm{n}_1)|^2]$, where
$|\psi(\bm{n}_2) / \psi(\bm{n}_1)|^2$ can be obtained from the network by
the above procedure, the sampling probability distribution of $\bm{n}$
becomes $|\psi(\bm{n})|^2 / \sum_{\bm{n}'} |\psi(\bm{n}')|^2 \equiv
P(\bm{n})$.
Using this sampling of $\bm{n}$, we can approximate as
\begin{equation} \label{sumPF}
\sum_{\bm{n}} P(\bm{n}) F(\bm{n}) \simeq \frac{1}{N_s} \sum_{i = 1}^{N_s}
F(\bm{n}_i) \equiv \langle F(\bm{n}) \rangle_M,
\end{equation}
if the number of samples $N_s$ is sufficient.
Using Eq.~(\ref{sumPF}), the expectation value of a quantity $\hat A$ is
calculated as
\begin{eqnarray} \label{A}
\langle \hat A \rangle & = & \frac{\sum_{\bm{n}, \bm{n}'} \psi^*(\bm{n})
\langle \bm{n} | \hat A | \bm{n}' \rangle \psi(\bm{n}')}
{\sum_{\bm{n}} |\psi(\bm{n})|^2} \nonumber \\
& = & \sum_{\bm{n}, \bm{n}'} P(\bm{n})
\langle \bm{n} | \hat A | \bm{n}' \rangle
\frac{\psi(\bm{n}')}{\psi(\bm{n})} \nonumber \\
& \simeq & \left\langle \sum_{\bm{n}'}
\langle \bm{n} | \hat A | \bm{n}' \rangle
\frac{\psi(\bm{n}')}{\psi(\bm{n})} \right\rangle_M \equiv
\langle \tilde A \rangle_M.
\end{eqnarray}

\subsection{Network optimization}
\label{s:opt}

The Hamiltonian for the system is given by
\begin{equation} \label{Hb}
\hat H = -J \sum_{\langle i, j \rangle} \hat{a}_i^\dagger \hat{a}_j
+ \frac{U}{2} \sum_i \hat{n}_i (\hat{n}_i - 1),
\end{equation}
where $J > 0$ is the hopping coefficient, and $U$ is the on-site
interaction energy.
The operator $\hat{a}_i$ annihilates a particle in the $i$th site, and
$\hat{n}_i = \hat{a}_i^\dagger \hat{a}_i$ is the number operator, where
the Bose commutation relation $[\hat{a}_i, \hat{a}_j^\dagger] =
\delta_{ij}$ is satisfied.

In order to optimize the network parameters $\bm{W}^{(n)}$ and
$\bm{b}^{(n)}$, we need to calculate the derivative of the expectation
value of the Hamiltonian $\langle \hat H \rangle$ with respect to these
network parameters.
Since the wave function $\psi(\bm{n})$ depends on the network parameters,
we have
\begin{eqnarray} \label{dHdw}
\frac{\partial \langle \hat H \rangle}{\partial w}
& = & \frac{\partial}{\partial w} 
\frac{\sum_{\bm{n}, \bm{n}'} \psi^*(\bm{n})
\langle \bm{n} | \hat H | \bm{n}' \rangle \psi(\bm{n}')}
{\sum_{\bm{n}} |\psi(\bm{n})|^2} \nonumber \\
& = & \frac{\sum_{\bm{n}, \bm{n}'} [ O_w^*(\bm{n}) + O_w(\bm{n}') ]
\psi^*(\bm{n}) \langle \bm{n} | \hat H | \bm{n}' \rangle \psi(\bm{n}')}
{\sum_{\bm{n}} |\psi(\bm{n})|^2} \nonumber \\
& & - \langle \hat H \rangle
\frac{\sum_{\bm{n}} [O_w^*(\bm{n}) + O_w(\bm{n})] |\psi(\bm{n})|^2}
{\sum_{\bm{n}} |\psi(\bm{n})|^2},
\end{eqnarray}
where $w$ is one of the network parameters and
\begin{equation}
O_w(\bm{n}) = \frac{1}{\psi(\bm{n})}
\frac{\partial \psi(\bm{n})}{\partial w}.
\end{equation}
The derivative $\partial\psi(\bm{n}) / \partial w$ can be calculated
systematically using the method of back propagation~\cite{Book}.
Using the stochastic approximation in Eqs.~(\ref{sumPF}) and (\ref{A}),
Eq.~(\ref{dHdw}) is obtained as
\begin{equation} \label{grad}
\frac{\partial \langle \hat H \rangle}{\partial w} \simeq
2 {\rm Re} \left( \langle O_w^* \tilde H \rangle_M
- \langle O_w^* \rangle_M \langle \tilde H \rangle_M \right).
\end{equation}

There are various ways to update the network parameters to reduce the
expectation value of the Hamiltonian.
In the steepest-descent method, the $i$th network parameter $w_i$
is updated as
\begin{equation} \label{SGD}
w_i \rightarrow w_i - \alpha \frac{\partial \langle \hat H \rangle}
{\partial w_i}, 
\end{equation}
where $\alpha < 1$ controls the magnitude of change in each update.
More efficient methods to update the network parameters have been
developed in the field of machine learning.
In the AdaGrad method, the network parameters are updated
as~\cite{Duchi}
\begin{eqnarray} \label{AdaGrad}
v_i & \rightarrow & v_i + \left( \frac{\partial \langle \hat H
\rangle}{\partial w_i} \right)^2, \nonumber \\
w_i & \rightarrow & w_i - \frac{\gamma}{\sqrt{v_i} + \epsilon}
\frac{\partial \langle \hat H \rangle}{\partial w_i}, 
\end{eqnarray}
where $\epsilon \ll 1$ avoids division by zero, and $\gamma < 1$.
The Adam method is given by~\cite{Kingma}
\begin{eqnarray} \label{Adam}
m_i & \rightarrow & \beta_1 m_i + (1 - \beta_1)
\frac{\partial \langle \hat H \rangle}{\partial w_i}, \nonumber \\
v_i & \rightarrow & \beta_2 v_i + (1 - \beta_2)
\left( \frac{\partial \langle \hat H \rangle}{\partial w_i} \right)^2,
\nonumber \\
w_i & \rightarrow & w_i - \delta \frac{m_i}{1 - \beta_1^\ell}
\frac{1}{\sqrt{\frac{v_i}{1 - \beta_2^\ell}} + \epsilon},
\end{eqnarray}
for the $\ell$th update, where $\beta_1 = 0.9$ and $\beta_2 = 0.999$ are
usually used, and $\delta < 1$.
The initial values of $v_i$ and $m_i$ in Eqs.~(\ref{AdaGrad}) and
(\ref{Adam}) are zero, and the parameters $\alpha$, $\gamma$, and $\delta$
are chosen so that the optimization works efficiently.
It is known that if a valley exists in the energy landscape, the
steepest-descent method may cause oscillation between the two sides of the
valley, while AdaGrad and Adam can avoid oscillation and efficiently
decrease the energy.
The stochastic reconfiguration method~\cite{Sorella} used in
Ref.~\citen{Carleo} may also be efficient, but it has a higher
computational cost for each update.
For the steepest descent, AdaGrad, and Adam, the cost for each update is
$O(N_w)$~\cite{Note}, where $N_w$ is the number of network parameters,
whereas the cost for the  stochastic reconfiguration method is $O(N_w^3)$.

The procedure for obtaining the approximate many-body ground state using
the ANN is as follows.
First, we initialize the network parameters $\bm{W}^{(n)}$ and
$\bm{b}^{(n)}$ with random numbers obeying normal distributions.
The standard deviations of the normal distributions are taken to be $1 /
\sqrt{N_{n - 1}}$ for $\bm{W}^{(n)}$ and $\bm{b}^{(n)}$~\cite{Xavier}.
The gradient in Eq.~(\ref{grad}) is calculated with Monte Carlo sampling
of typically $N_s = 1000$ samples.
The network parameters are then updated using the steepest-descent method
in Eq.~(\ref{SGD}), AdaGrad in Eq.~(\ref{AdaGrad}), or Adam in
Eq.~(\ref{Adam}).
The procedures of Monte Carlo sampling and parameter updating are
repeated until the energy $\langle \tilde H \rangle_M$ converges.

\section{Numerical Results}
\label{s:bose}

We consider a one-dimensional (1D) system of $N$ bosons on $M$ sites with
the periodic boundary condition.
The ground state of the Hamiltonian in Eq.~(\ref{Hb}) for $U = 0$ is
$(\hat{b}^\dagger)^N / \sqrt{N!} |\bm{0} \rangle$, where $\hat{b} = \sum_i
\hat{a}_i / \sqrt{M}$ and $|\bm{0} \rangle$ is the vacuum.
The ground-state energy for $U = 0$ is $E = -2 N J$.

\begin{figure}[tb]
\begin{center}
\includegraphics[width=8.5cm]{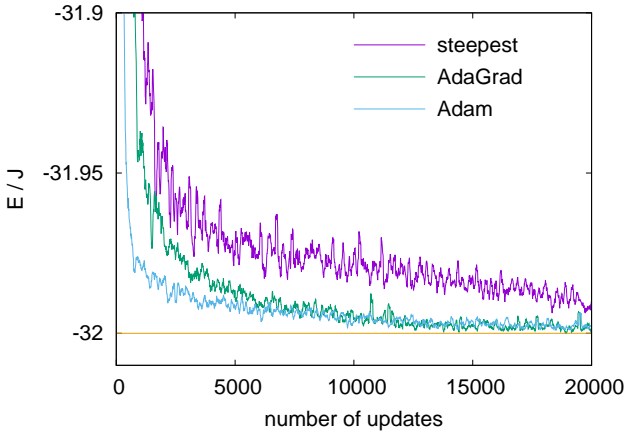}
\end{center}
\caption{
Convergence of the ground-state energy as a function of the number of
updates of the network parameters for the steepest-descent, AdaGrad, and
Adam methods in Eqs.~(\ref{SGD}), (\ref{AdaGrad}), and (\ref{Adam}),
respectively.
The 1D Bose-Hubbard model with $N = 16$ particles in $M = 16$ sites for
$U_B / J = 0$.
The update rates $\alpha = 0.3$, $\gamma = 0.3$, and $\delta = 0.005$ are
chosen to optimize the convergence.
The ANN is a fully-connected network having a single hidden layer with
$N_1 = 40$ units.
The average energy of the previous 100 updates is shown.
The horizontal line indicates the exact ground-state energy.
}
\label{f:converge}
\end{figure}
We first compare three optimization methods: steepest descent, AdaGrad,
and Adam.
Figure~\ref{f:converge} shows the energy $\langle \tilde H \rangle_M$ as a
function of the number of updates using the three methods.
We can see that the energy decreases quickly for the initial $\sim 1000$
updates and then gradually converges to the final value.
The convergence is faster for AdaGrad and Adam than for the
steepest-descent method.
The update rates used in Fig.~\ref{f:converge} are $\alpha = 0.3$, $\gamma
= 0.3$, and $\delta = 0.005$, which are chosen so that the energy
convergence becomes the most efficient.
When these parameters are smaller, the convergence becomes slower, and
when the parameters are too large, the calculation becomes unstable.
In the following calculations, we adopt Adam with $\delta = 0.005$ for the
network updates.

\begin{figure}[tb]
\begin{center}
\includegraphics[width=8.5cm]{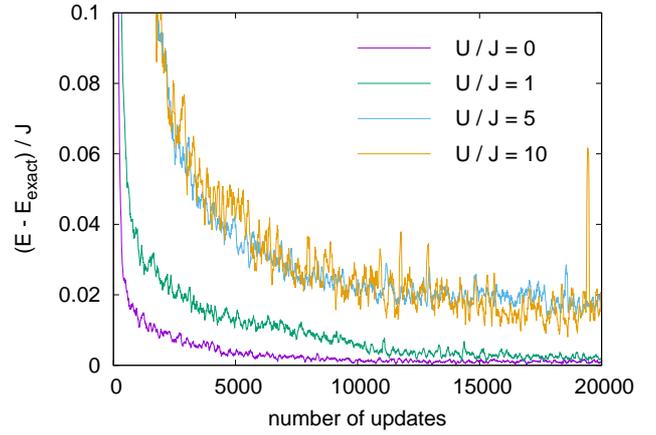}
\end{center}
\caption{
Dependence of the energy convergence on the on-site energy $U$ for the 1D
Bose-Hubbard model with $N = 14$ and $M = 14$.
The exact energy $E_{\rm exact}$ is obtained by the exact diagonalization
of the Hamiltonian.
The ANN is a fully-connected network having a single hidden layer with
$N_1 = 40$ units.
The average energy of the previous 100 updates is shown.
}
\label{f:udep}
\end{figure}
Figure~\ref{f:udep} shows the $U$-dependence of the energy convergence
with respect to the network updates.
For the noninteracting case, the deviation from the exact energy rapidly
converges to $(E - E_{\rm exact}) / J \lesssim 0.002$, whereas for a large
on-site interaction $U$, the convergence is slow and $(E - E_{\rm exact})
/ J \sim 0.02$ even after 20000 updates.
Slow convergence occurs for $U / J \gtrsim 5$, which is the Mott insulator
regime.
Increasing the number of updates to 60000, the energy difference becomes 
$(E - E_{\rm exact}) / J \simeq 0.01$ for $U / J = 10$.

\begin{figure}[tb]
\begin{center}
\includegraphics[width=8.5cm]{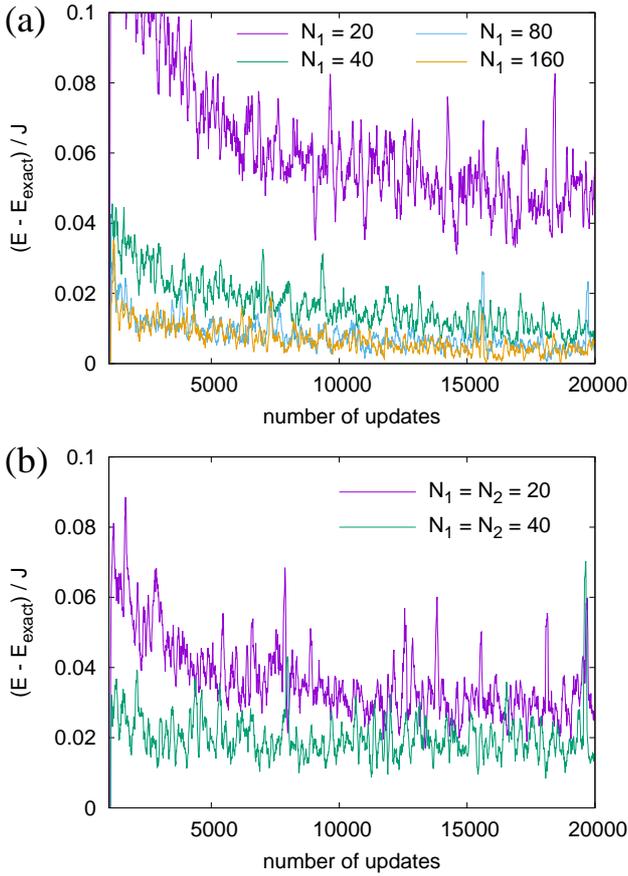}
\end{center}
\caption{
Dependence of energy convergence on the size and depth of the network
for the 1D Bose-Hubbard model with $N = 14$, $M = 14$, and $U / J = 10$.
(a) Fully-connected network with a single hidden layer.
The number of units in the hidden layer is $N_1 = 20$, 40, 80, and 160.
(b) Fully-connected network with two hidden layers.
The numbers of units in the hidden layers are $N_1 = N_2 = 20$ and
$N_1 = N_2 = 40$.
The value of $U$ is linearly ramped from $0$ to $10J$ in the first 1000
updates.
The exact energy $E_{\rm exact}$ is obtained by the exact diagonalization
of the Hamiltonian.
The average energy of the previous 100 updates is shown.
}
\label{f:layer}
\end{figure}
In order to improve the slow convergence for large $U$, we change the size
and depth of the fully-connected network.
Figure~\ref{f:layer}(a) shows the dependence of the energy convergence on
the number of hidden units $N_1$ for a single hidden layer.
The energy convergence is improved as $N_1$ is increased, and saturates at
$N_1 \gtrsim 100$.
Figure~\ref{f:layer}(b) shows the case of a fully-connected network with
two hidden layers.
Although the convergence for $N_1 = N_2 = 20$ in Fig.~\ref{f:layer}(b) is
better than that of the single hidden layer with $N_1 = 20$ in
Fig.~\ref{f:layer}(a), the convergence for $N_1 = N_2 = 40$ becomes
worse compared with that of the single hidden layer with $N_1 = 40$.
This implies that the increase in the depth of the fully-connected network
may improve the energy convergence due to the increase in the capability
of the network, but it can also be counterproductive due to the increase
in the complexity of the network.

In Fig.~\ref{f:layer}, the value of $U$ is linearly increased from 0 to
$10J$ in the initial 1000 updates.
Such a gradual ramp of $U$ prevents the network from becoming trapped in a
local minimum of the energy.
Furthermore, the initial ramp of $U$ accelerates the energy convergence
(compare the $N_1 = 40$ line in Fig.~\ref{f:layer}(a) with the $U / J =
10$ line in Fig.~\ref{f:udep}).
In the following calculations, we linearly ramp the value of $U$ in the
initial 1000 updates.

\begin{figure}[tb]
\begin{center}
\includegraphics[width=8.5cm]{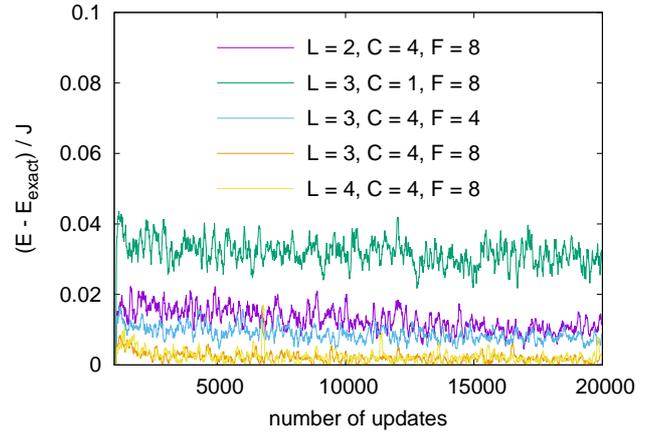}
\end{center}
\caption{
Energy convergence for the convolutional neural networks.
The 1D Bose-Hubbard model with $N = 14$, $M = 14$, and $U / J = 10$.
The input layer is followed by the $(L - 1)$ convolution layers with
filter size $F \equiv F_1 = \cdots = F_{L-1}$ and number of output
channels $C \equiv C_1 = \cdots = C_{L-1}$.
The final hidden layer is fully connected to the output layer.
The value of $U$ is linearly ramped from $0$ to $10J$ in the first 1000
updates.
The exact energy $E_{\rm exact}$ is obtained by the exact diagonalization
of the Hamiltonian.
The average energy of previous 100 updates is shown.
}
\label{f:conv}
\end{figure}
We next consider the convolutional neural network in
Eqs.~(\ref{conv1})-(\ref{conv3}).
The input layer is connected to the first convolution layer through
filters of size $F_1 = F$, which produces $C_1 = C$ channels; i.e., 
$N_0 = M$ input units are connected to $N_1 = C M$ hidden units.
When $L = 2$ (a single convolution layer), $N_1 = C M$ units produced
by the convolution layer are fully connected to the output units.
When $L > 2$ ($L - 1$ consecutive convolution layers), $C M$ output
units of the convolution layer are input into the next convolution layer,
which also produces $C M$ output units, and finally the $(L - 1)$th
layer is fully connected to the output layer. 
Comparing $(L, C, F) = (2, 4, 8)$, $(3, 4, 8)$, and $(4, 4, 8)$ in
Fig.~\ref{f:conv}, we find that a network with two convolution layers
($L = 3$) exhibits much better convergence than a network with a single
convolution layer ($L = 2$).
However, a further increase in the convolution layer ($L = 4$) results in
no improvement, and therefore the improvement saturates at $L = 3$.
The decrease in the number of channels $(C = 1)$ or the size of the
filters $(F = 4)$ makes the convergence worse, while no improvement is
obtained for $C > 4$ and $F > 8$.
Thus, in the present case, the convolutional ANN with $(L, C, F) = (3, 4,
8)$ yields the fastest convergence and the smallest energy, which is much
better than the fully-connected ANNs.

\begin{figure}[tb]
\begin{center}
\includegraphics[width=8.5cm]{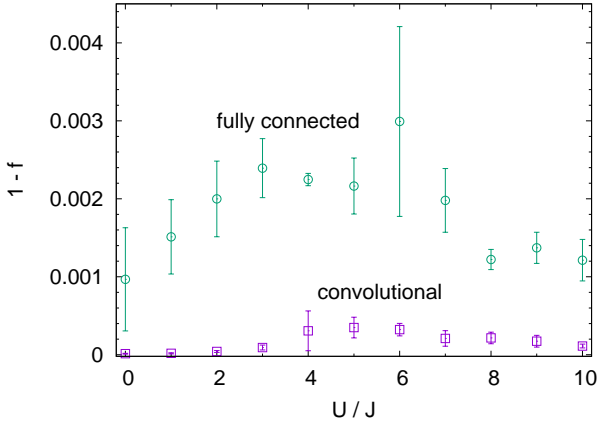}
\end{center}
\caption{
Error in the many-body wave function.
The 1D Bose-Hubbard model with $N = 14$ and $M = 14$.
The fidelity $f$ is defined in Eq.~(\ref{f}).
The fully-connected network (circles) has a single hidden layer with $N_1
= 40$ units.
The convolutional network (squares) has two convolution layers with filter
size $F_1 = F_2 = 8$ and $C_1 = C_2 = 4$ channels.
The fidelity is calculated after 10000 updates.
The error bars represent the standard deviation of five results with
different random numbers in the initial network parameters.
}
\label{f:fidelity}
\end{figure}
We evaluate the accuracy of the ground-state wave function obtained by the
optimized network.
The fidelity of the wave function is defined as
\begin{equation} \label{f}
f = \frac{\left|\sum_{\bm{n}} \psi^*(\bm{n}) \psi_{\rm exact}(\bm{n}) 
\right|^2}{\sum_{\bm{n}} |\psi(\bm{n})|^2},
\end{equation}
where $\psi$ is the wave function generated by the ANN, and
$\psi_{\rm exact}$ is that obtained by the exact diagonalization of the
Hamiltonian.
The sum in Eq.~(\ref{f}) is taken for all possible $\bm{n}$.
When the wave function $\psi$ is the exact ground state, the fidelity in
Eq.~(\ref{f}) is $f = 1$.
Figure~\ref{f:fidelity} shows the error $1 - f$ as a function of $U / J$.
In order to obtain each value in Fig.~\ref{f:fidelity}, 10000 updates of
the network parameters are performed, and the results of five runs with
different initial network parameters produced by random numbers are
averaged, where error bars represent standard deviation.
We find that the errors $1 - f$ are smaller for the convolutional neural
network than for the fully-connected network, which is consistent with the
results shown in Figs.~\ref{f:layer} and \ref{f:conv}.

The results in Figs.~\ref{f:converge}-\ref{f:fidelity} indicate that the
quantum many-body state is stored in the ANN very efficiently.
According to Eq.~(\ref{nfock}), the number of Fock-state bases needed for
expressing the exact wave function is $N_{\rm Fock} = 20058300$ for $N = M
= 14$.
On the other hand, the number of network parameters for the
fully-connected network used in Fig.~\ref{f:fidelity} is $N_{\rm FC} =
680$ for $N_1 = 40$ (Eq.~(\ref{NFC})), and that for the convolutional
network is $N_{\rm conv} = 280$ for $F = 8$ and $C = 4$
(Eq.~(\ref{NConv})), which are much smaller than $N_{\rm Fock}$.
Therefore, the information of the many-body wave function is compressed
and stored in the ANN very efficiently.
It is remarkable that such compressibility of the wave function is
automatically achieved in the optimization process of the network.

\begin{figure}[tb]
\begin{center}
\includegraphics[width=7cm]{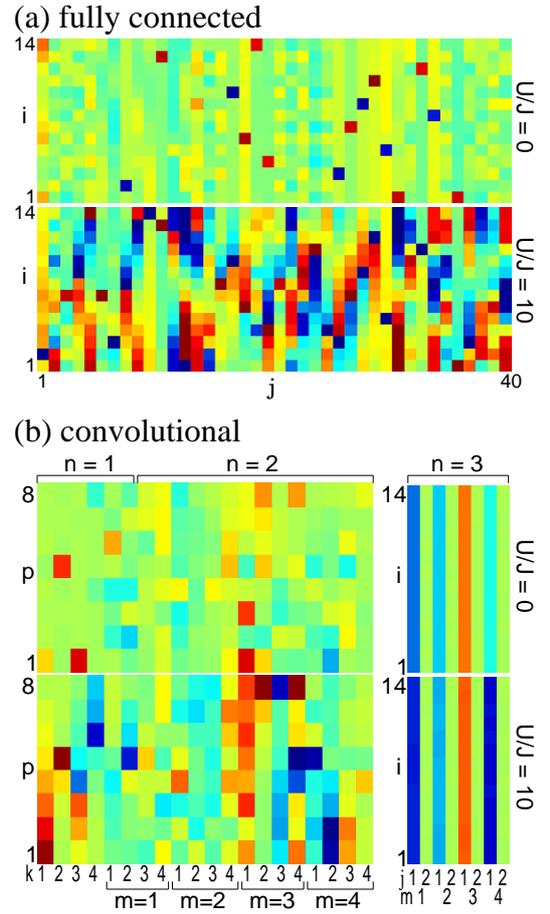}
\end{center}
\caption{
Weights of optimized networks for the 1D Bose-Hubbard model with $N = 14$
and $M = 14$.
(a) Weights $W_{ij}^{(1)}$ of the fully-connected neural network having a
single hidden layer with $N_1 = 40$, where $1 \leq i \leq M$ and $1 \leq j
\leq N_1$.
(b) Weights $W_{pk}^{(1)}$, $W_{pmk}^{(2)}$, and $W_{mij}^{(3)}$ for the
convolutional neural network having two convolutional layers with filter
size $F = 8$ and $C = 4$ channels, and a fully-connected layer, where $1
\leq p \leq F$, $1 \leq k \leq C$, $1 \leq m \leq C$, $1 \leq i \leq M$,
and $j = 1, 2$.
}
\label{f:weight}
\end{figure}
To see how the quantum many-body states are stored in the ANNs, the
network weights $\bm{W}^{(n)}$ after 10000 updates are visualized in
Fig.~\ref{f:weight}.
The fluctuations in the weights are larger for $U / J = 10$ than for $U /
J = 0$.
For the fully-connected network in Fig.~\ref{f:weight}(a), no regularity
or meaningful pattern appears in the weights.
Although the many-body wave function produced by the network has
translational symmetry, the weights have no apparent translational
symmetry with respect to the site index $i$ for the fully-connected
network in Fig.~\ref{f:weight}(a).
By contrast, for the convolutional network in Fig.~\ref{f:weight}(b), the
weights $\bm{W}^{(3)}$ in the final fully-connected layer are independent
of the site index $i$.
This indicates that the network has translational symmetry, since the
convolution layers have translational symmetry by the definition
(i.e., the filters are shared by all of the site indices $j$ in
Eqs.~(\ref{conv1}) and (\ref{conv2})).

The translational symmetry of the convolutional network may be related to
the faster convergence and smaller energy achieved in the optimization
process, as compared with the fully-connected ANN.
Let us consider the capability of both ANNs.
For example, the convolutional ANN with $(L, C, F) = (3, 4, 8)$ in
Fig.~\ref{f:conv} is a subset of the fully-connected ANN having two hidden
layers with $N_1 = N_2 = M C = 56$, namely, the former is realized by the
latter with constraints on $\bm{W}^{(n)}$ and $\bm{b}^{(n)}$.
Therefore, potentially, the latter has the ability to represent the
many-body wave function more accurately than the former.
Nevertheless, the fully-connected ANN is worse than the convolutional ANN,
because the large degree of freedom of the fully-connected ANN makes its
optimization inefficient.
The convolutional ANN, on the other hand, takes into account the
translational symmetry of the system, and a smaller number of network
parameters makes the optimization efficient.
The filters may also be suitable for capturing the local correlations
produced by the on-site local interaction.
Thus, presumably, the convolutional ANN is quite compatible with
physical systems with local interaction and translational symmetry.

\begin{figure}[tb]
\begin{center}
\includegraphics[width=8.5cm]{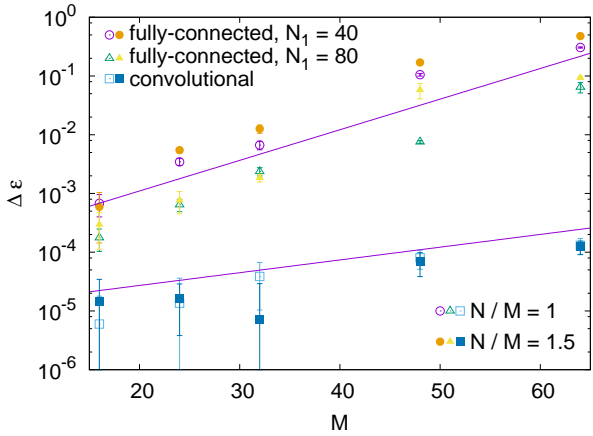}
\end{center}
\caption{
Energy difference between the present method and the DMRG as a function of
the number of sites $M$ for $N / M = 1$ (open plots) and $N / M = 1.5$
(filled plots) with $U / J = 1$.
The ANNs are the fully-connected neural networks having a single hidden
layer with $N_1 = 40$ (circles) and $N_1 = 80$ (triangles) units, and a
convolutional neural network having two convolutional layers with filter
size $F_1 = F_2 = 8$ and  $C_1 = C_2 = 4$ channels (squares).
The lines are proportional to $\exp(0.12 M)$ and $\exp(0.05 M)$.
The energy is calculated from 10000 samples after 10000 updates.
The error bars represent the standard deviation of five results with
different random numbers in the initial network parameters.
}
\label{f:large}
\end{figure}
We next consider larger systems for which exact diagonalization is
difficult or almost impossible.
Instead of exact diagonalization, we adopt the method of the density
matrix renormalization group (DMRG)~\cite{White,ALPS}, which is known to
be very accurate for 1D many-body problems.
Figure~\ref{f:large} shows the difference between the energy per particle
obtained by the present ANN method, $E / N$, and that by the DMRG,
$E_{\rm DMRG} / N$,
\begin{equation}
\Delta \varepsilon = \frac{1}{N} (E - E_{\rm DMRG}),
\end{equation}
as a function of the number of sites $M$.
For the fully-connected neural network, the error $\Delta \varepsilon$
appears to increase exponentially with $M$ as $\Delta \varepsilon \sim
\exp(\kappa M)$ with $\kappa \simeq 0.12$.
When the number of hidden units is increased from $N_1 = 40$ to 80,
the accuracy is improved by an order of magnitude.
The convolutional neural network yields energies with much better
precision.
The error $\Delta \varepsilon$ also appears to increase with $M$ as
$\Delta \varepsilon \sim \exp(\kappa M)$, where $\kappa \simeq 0.05$ is
smaller than that of the fully-connected neural network.
The results for larger particle density $N / M = 1.5$ are similar to those
for $N / M = 1$.

\begin{figure}[tb]
\begin{center}
\includegraphics[width=8.5cm]{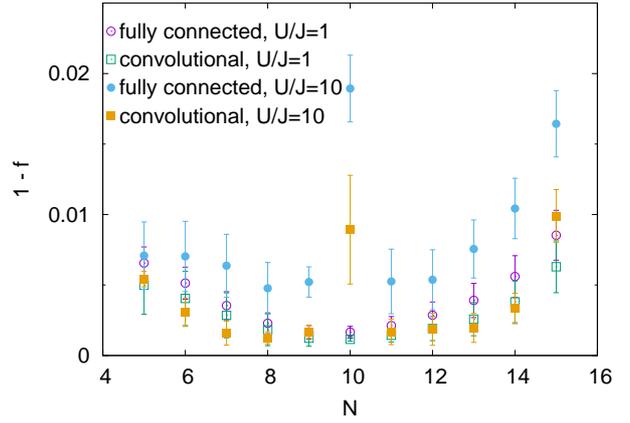}
\end{center}
\caption{
A single network is optimized for different numbers of atoms $N = 5$, 6,
$\cdots$, 15.
The 1D Bose-Hubbard model with $M = 10$ and $U / J = 1$ and 10.
The fully-connected network has a single hidden layer with $N_1 = 40$
units.
The convolutional network has two convolution layers with filter size $F_1
= F_2 = 8$ and  $C_1 = C_2 = 4$ channels.
In total, 10000 updates are made with randomly chosen $5 \leq N \leq 15$
in each update.
The fidelity is defined in Eq.~(\ref{f}).
The error bars represent the standard deviation of five results with
different random numbers in the initial network parameters.
}
\label{f:multin}
\end{figure}
We have thus far considered the case in which a single many-body ground
state is stored in the ANN.
We next try to obtain the ground states for different numbers of atoms $N$
by a single optimized ANN.
In order to optimize the network in such a manner, in each update step
described in Sec.~\ref{s:opt}, we choose $N$ randomly, and the Metropolis
samplings are performed for $\bm{n}$ with $N$ particles.
In Fig.~\ref{f:multin}, we randomly choose $N$ from 5 to 15 in each update
step, and a total of 10000 updates are performed.
The fidelity is then calculated for each $N$ using the optimized network.
Figure~\ref{f:multin} shows that $1 - f$ is always smaller than 0.01
for $U / J = 1$, which indicates that the multiple many-body ground states
are stored in the single ANN.
However, the precision of each state is worse than in the case in which
the ANN is optimized for a specific $N$ (see Fig.~\ref{f:fidelity}).
For $U / J = 10$, the error is prominent at $N = 10$, which is the Mott
insulator state with unit filling.
The convolutional network is better than the fully-connected network also
in this case.

\begin{figure}[tb]
\begin{center}
\includegraphics[width=8.5cm]{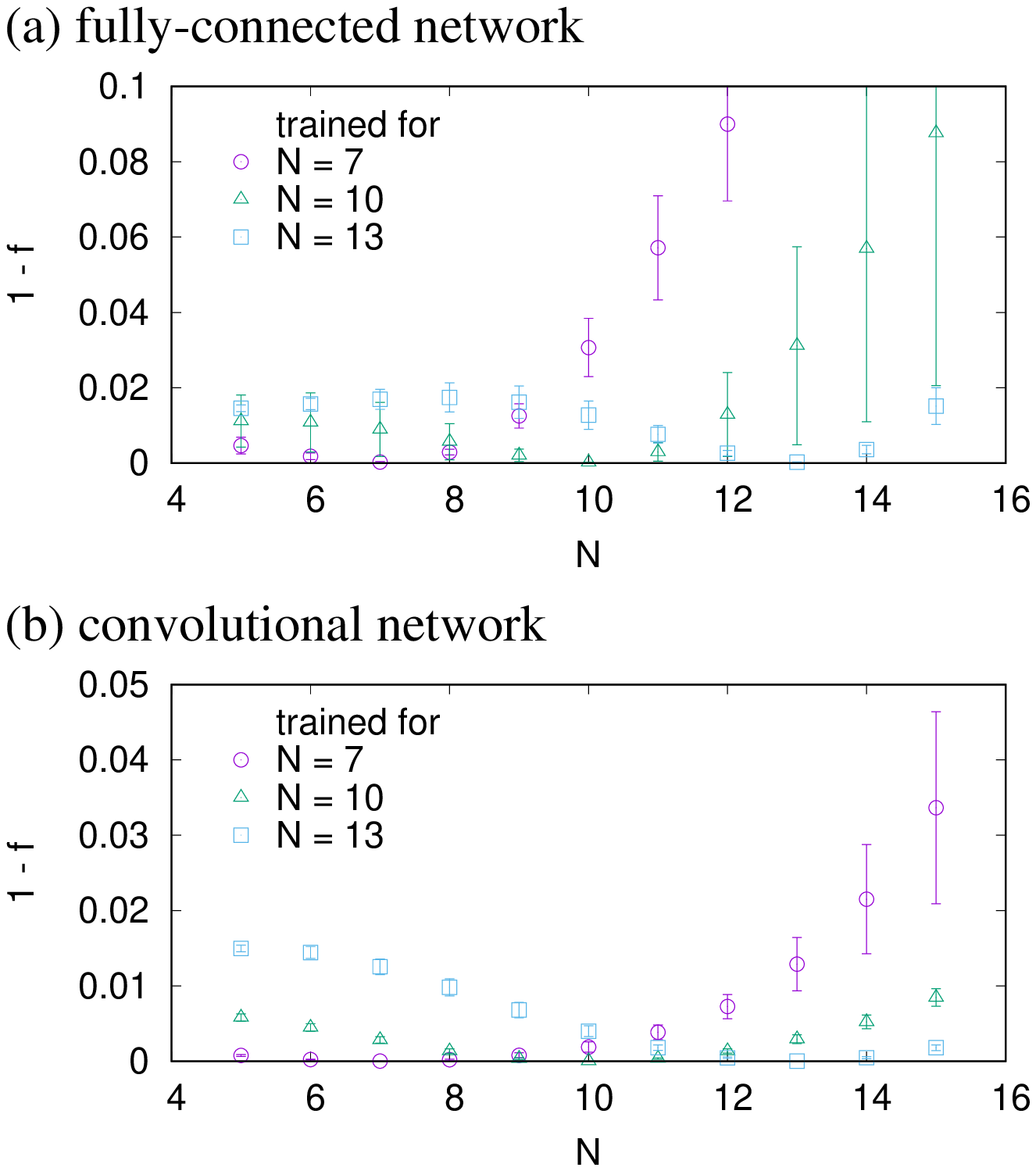}
\end{center}
\caption{
Many-body wave functions for $N = 5$, 6, $\cdots$, 15 generated by a
single ANN that is only optimized for $N = 7$, 10, or 13.
The 1D Bose-Hubbard model with $M = 10$ and $U / J = 1$ and 10.
(a) Fully-connected network having a single hidden layer with $N_1 =
40$ units.
(b) Convolutional network having two convolution layers with a filter
size of $F_1 = F_2 = 8$ and  $C_1 = C_2 = 4$ channels.
The fidelity is defined in Eq.~(\ref{f}).
The error bars represent the standard deviation of five results with
different random numbers in the initial network parameters.
}
\label{f:othern}
\end{figure}
Figure~\ref{f:othern} shows similar results, but the network is optimized
only for a specific $N$ ($= 7$, 10, or 13).
For example, when the network is optimized for $N = 7$, the fidelity is
best at $N = 7$, as expected.
Note that the fidelity is good not only for $N = 7$, but also for $N = 6$
and 8.
For the convolutional network optimized for $N = 10$, the fidelity is
below 0.01 in the range of $N$ shown in Fig.~\ref{f:othern}(b).
These results imply that the present method may also be used for
extrapolating (or interpolating) the quantum many-body states, i.e., the
ANN optimized for certain parameters may generate approximate many-body
states for other parameters.

\section{Conclusions}
\label{s:conc}

In conclusion, we have developed a method to obtain the quantum many-body
ground state of the Bose-Hubbard model using a feedforward artificial
neural network.
Although the simple steepest-descent method was only employed in the
previous Letter~\cite{Letter}, we examined AdaGrad and Adam in the present
paper and found that the convergence is better (Fig~\ref{f:converge}).
The accuracy of the present method becomes worse as the on-site
interaction $U$ is increased (Fig.~\ref{f:udep}).
In order to increase the accuracy, we investigated a deep (multi-layer)
fully-connected network.
However, a single hidden layer with a sufficient number of hidden units is
found to be better than multiple hidden layers (Fig.~\ref{f:layer}).
We then investigated the deep convolutional network and found it to be
much more efficient than the fully-connected network (Figs.~\ref{f:conv}
and \ref{f:fidelity}).
We found that the convolutional network has translational symmetry
(Fig.~\ref{f:weight}).
The convolutional network is also promising for studying large systems
(Fig.~\ref{f:large}).
Multiple quantum many-body states can be stored in a single ANN
(Figs.~\ref{f:multin} and \ref{f:othern}).

At present, it is unclear whether the present method of obtaining
quantum many-body states can surpass other existing methods in terms of
precision and computational resources.
For 1D cases, the DMRG seems better, whereas for 2D and 3D cases, the
present method may have advantages.
At the very least, the ANN is a very versatile scheme for representing
quantum many-body states and may provide an initial choice to tackle
quantum many-body problems for which effective solution methods are
unknown.
In order to confirm this possibility, we must confirm that the method
works for various other quantum many-body problems.

\begin{acknowledgments}
The present study was supported by JSPS KAKENHI Grant Numbers JP16K05505,
JP17K05595, JP17K05596, and JP25103007.
\end{acknowledgments}


\begin{thebibliography}{9}
\bibitem{Note}
In the calculations performed in the present paper, it takes a few minutes
for 10000 updates using my work station (Intel Xeon E5-2697A v4).
\bibitem{ALPS}
The DMRG calculations are performed using the ALPS code.
\end{thebibliography}
\end{document}